\documentclass{article}
\usepackage{arxiv}
\usepackage[utf8]{inputenc} 
\usepackage[T1]{fontenc}    
\usepackage{hyperref}       
\usepackage{url}            
\usepackage{booktabs}       
\usepackage{amsfonts}       
\usepackage{nicefrac}       
\usepackage{microtype}      
\usepackage{lipsum}		
\usepackage{graphicx}
\usepackage{xcolor}

\begin{document}

\title{Tropes in films: an initial analysis}

\author{Rubén Héctor García-Ortega\thanks{Department of Computer Architecture and Technology, University of Granada} \and Pablo García-Sánchez\thanks{Department of Languages and Computer Systems, University of Granada} \and JJ Merelo\thanks{Department of Computer Architecture and Technology, University of Granada}}

\maketitle

\begin{abstract}
    TVTropes is a wiki that describes tropes and which ones are used in which artistic work.
    We are mostly interested in films, so
    after releasing the TropeScraper Python module that extracts data from this site,
    in this report we use scraped information
    to describe statistically how tropes and films are related to each other
    and how these relations evolve in time.
    In order to do so, we generated a dataset through the tool TropeScraper in April 2020.
    We have compared it to the latest snapshot of DB Tropes,
    a dataset covering the same site and published in July 2016,
    providing descriptive analysis, studying the fundamental differences
    and addressing the evolution of the wiki in terms of the number of tropes,
    the number of films and connections.
    The results show that the number of tropes and films doubled their value and quadrupled their relations,
    and films are, at large, better described in terms of tropes.
    However, while the types of films with the most tropes has not changed significantly in years,
    the list of most popular tropes has.
    This outcome can help on shedding some light on how popular tropes evolve,
    which ones become more popular or fade away, and in general how a set of tropes represents a film and
    might be a key to its success.
    The dataset generated, the information extracted, and the summaries provided
    are useful resources for any research involving films and tropes.
    They can provide proper context and explanations about the behaviour
    of models built on top of the dataset, including the generation of new content or its use in machine learning.
\end{abstract}


\section{Motivations} \label{sec:problem}

TV Tropes~\cite{tvtropes} is an outstanding source of knowledge because it describes thousands of tropes
and relates them to artistic works, providing examples with a broad context in terms of the characters,
places and actions.
These relations, once extracted and processed, have shown to be very valuable in the field of
Artificial Intelligence, in both content generation and quality prediction, as in ~\cite{doi:10.1111/exsy.12525}.

Since TV Tropes is a non-structured wiki-style database, most researchers have accessed to its data
through a structured n-tuple-based database called DB Tropes~\cite{DBTropes32:online}, which we enriched and analyzed in \cite{garcia2018overview}. Unfortunately, BDTropes remains unattended since July 2016, while the prolific community of TV Tropes
has doubled the number of films and their tropes since then.
For this reason, in order to allow the researchers to use the films and tropes from TV Tropes,
we implemented a scraper called TropeScraper~\cite{tropescr46:online} and licensed it as free software,
so it becomes part of the Python ecosystem.
This new tool, which takes days to download the whole database, has allowed us, and now you,
to make a more up-to-date analysis of the tropes used in films. The rest of the paper will be devoted to this.

Although the quantity of the information extracted is promising, doubling the number of films and tropes in TV Tropes,
an exploratory statistical analysis shows that the data is still incomplete, with distributions
that point out to huge biases due to the popularity of the films and tropes.
The detection of these biases is essential to explain the outcomes of future automatic learning models
build on top of the dataset, for example, according to~\cite{doi:10.1111/exsy.12525}
the number of tropes in TV Tropes is proportional to the popularity of a movie and inversely proportional to its age.

In the following report, we evaluate the dataset of films and tropes from TV Tropes as extracted April 2020,
because knowing their absolute numbers, frequencies and the most relevant ones can be critical
in different stages of future researches in automatic authoring tools,
from data wrangling to technique selection, configuration and outcome interpretation.
As a secondary goal, we compare and analyse how TV Tropes has evolved from July 2016 to April 2020
in order to detect general trends that have emerged which could help on the prediction of TV Tropes
in some years away.
Finally, since TV Tropes is an unstructured wiki and we have to follow different heuristics
to crawl the pages correctly, we also assess the quality of the scraping process in this report,
considering possible inconsistencies and improvements based on the results.

We divide the rest of the report as follows:
Section~\ref{sec:acquisition} explains the data acquisition phase.
The analysis in Sections~\ref{sec:analysys1} and~\ref{sec:analysys2} includes the two perspectives of
the relation between films and tropes; in other words, how many tropes appear in each movie and vice versa.
The conclusion in Section~\ref{sec:conclusions} highlights and summarises the outcomes.

\section{Notes on data Acquisition} \label{sec:acquisition}

It is important to remark that we have improved the original scraper that we used in ~\cite{doi:10.1111/exsy.12525}.
The first stable version of the scraper (v1.0.2) crawled through the film pages and scraped  tropes found in every film.
The data seemed to correlate with DBTropes; however, the analysis of the top tropes showed missing information because
it was not considering tropes that were very common in 2020.
After studying the situation, we discovered
that counting the tropes listed in the film pages was not enough;
the trope pages also include the list of films that they belong to,
and this information complements the previous one, increasing the relations between films and tropes.
Finally, some massively used tropes divide their occurrences into wiki pages
that are in the same level of hierarchy than films and tropes,
although semantically they are at the trope level.
For these reasons, we evolved the scraper to find the relations from the film's perspective,
from the trope's perspective and also, considering pagination,
providing a more solid dataset that is very aligned to the data in 2016
and the changes stated inside TV Tropes since then.
We published it as version v1.1.0.
As a side effect, its scraping time passed from a couple of hours to days,
because it included ten times more pages to process.

\section{Analysis of the number of tropes used in the films} \label{sec:analysys1}

The descriptive analysis of the films regarding their tropes (Table~\ref{tab:decriptive_analysis_tropes_by_film})
reveals the growth of the number of films by 99.6\%.
The average number of tropes per film has also grown by 279.38\%.
In other words, the dataset doubles the number of films (2x) and quadruples the number of tropes by film (4x).
It is interesting to point out that the scraper has been able to find five movies without tropes.
However, in all the cases, we could manually check and confirm that the films do not have a page in TV Tropes,
so they are listed in the categories by mistake.

\begin{figure}[htpb]
\center

\includegraphics[width= 10cm]{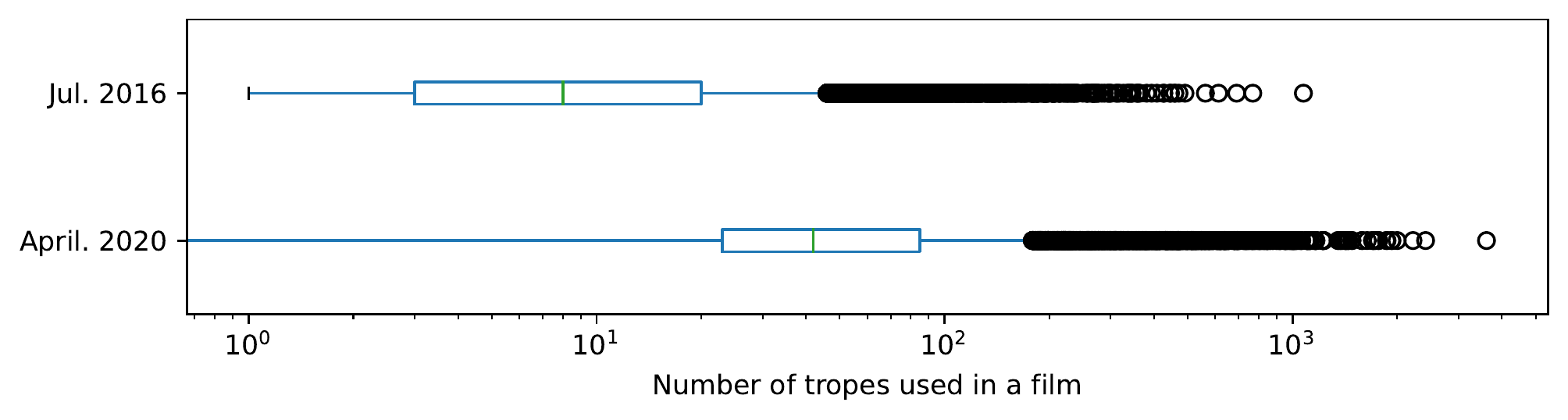}

\caption{Comparative boxplots of trope occurrences by film between July 2016 and April 2019}
\label{fig:boxplot_tropes_by_film}
\end{figure}

The increase of the frequency is visible through the comparative boxplots (Figure~\ref{fig:boxplot_tropes_by_film}),
where a reduced interquartile range shows that
there are fewer differences among the number of tropes in 50\% of the films,
and moved to the top, that shows how the average number of tropes per film has grown.

\begin{table}[htpb]
\center

\begin{tabular}{lrr}
\toprule
{} &  Jul. 2016 &  April. 2020 \\
\midrule
min      &      1.000 &        0.000 \\
max      &   1075.000 &     3611.000 \\
nobs     &   6296.000 &    12567.000 \\
mean     &     21.984 &       83.402 \\
kurtosis &     85.792 &       76.165 \\
skewness &      7.046 &        6.750 \\
variance &   2164.167 &    20937.037 \\
\bottomrule
\end{tabular}

\caption{Comparative Descriptive analysis of trope occurrences by film between July 2016 and April 2019}
\label{tab:decriptive_analysis_tropes_by_film}
\end{table}

\begin{figure}[htpb]
\center

\includegraphics[width= 10cm]{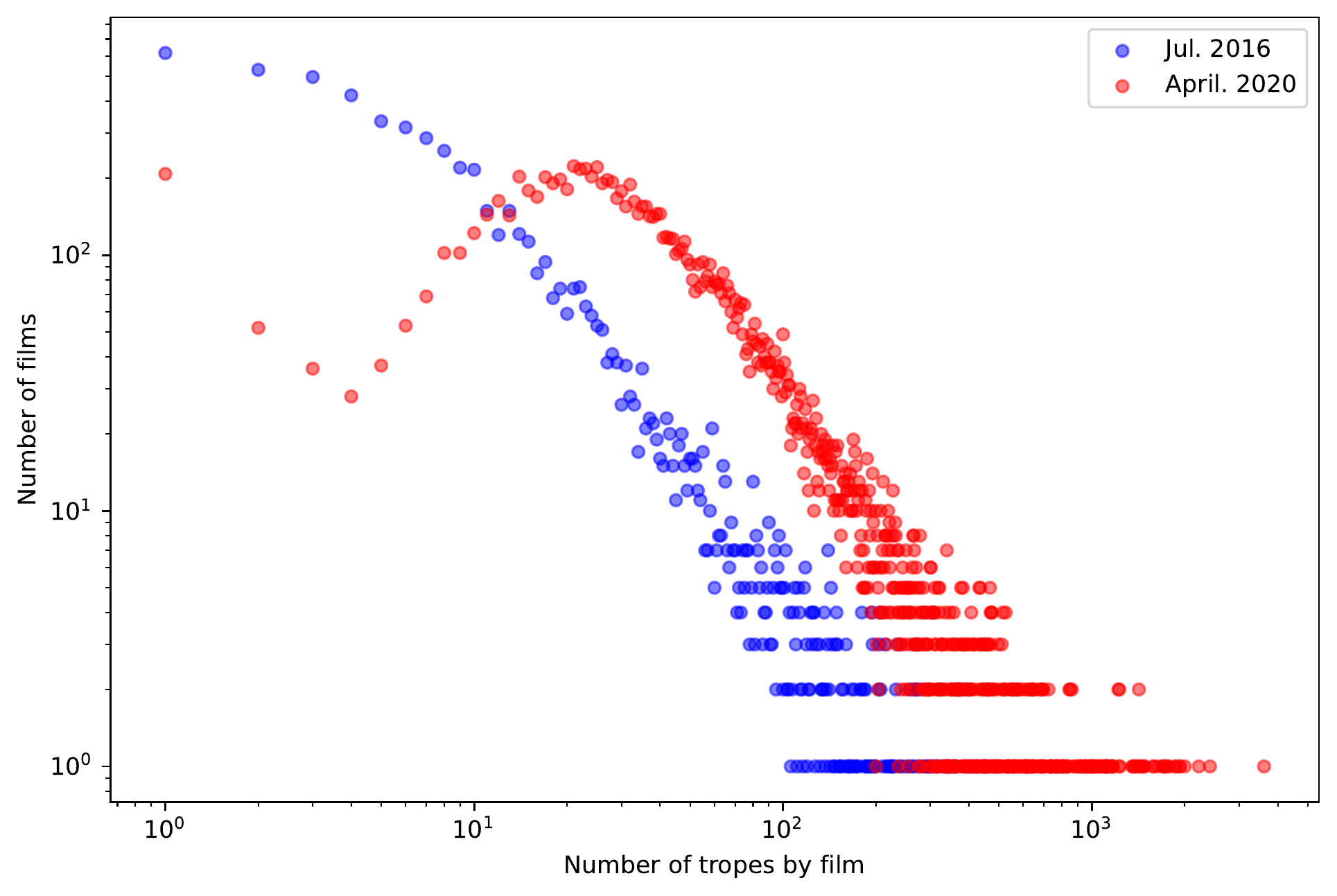}

\caption{Comparative boxplots of film occurrences by tropes between July 2016 and April 2019}
\label{fig:frequency_tropes_by_film}
\end{figure}

If we represent the number of tropes per film concerning the number
of movies (Figure~\ref{fig:frequency_tropes_by_film}),
we can observe a clear trend: the number of films with just one trope is high,
and the number of films decreases as the number of tropes increases, reaching the minimum in 4 tropes.
Then, the value reaches a maximum value close to 20 tropes and decreases again
following a long-tail distribution.

The films with more tropes (Table~\ref{tab:top_films_by_number_of_tropes})
are adventure/action film franchises, such as James Bond, The Avengers and Star Wars.
If we compare the films in 2020 and 2016, we can observe that, in general, the top ones remain.
As a rule, the films that were previously at the top have multiplied the number of tropes by up to 10.

\begin{table}[htpb]
\center

\begin{tabular}{llrlrl}
\toprule
{} &                                 Film (Jul. 2016) & Tropes &                               Film (April. 2020) & Tropes &    Increment \\
\midrule
0  &                      \textcolor{blue}{JamesBond} &   1075 &                      \textcolor{blue}{JamesBond} &   3611 &     +235.9\% \\
1  &              \textcolor{blue}{TheLordOfTheRings} &    769 &              \textcolor{blue}{TheLordOfTheRings} &   2413 &     +213.8\% \\
2  &                      \textcolor{blue}{TheMatrix} &    691 &                \textcolor{blue}{TheAvengers2012} &   2220 &     +558.8\% \\
3  &                  \textcolor{blue}{TheDarkKnight} &    613 &                  \textcolor{blue}{TheDarkKnight} &   1999 &     +226.1\% \\
4  &                       \textcolor{blue}{StarTrek} &    562 &                                         ANewHope &   1932 &     +732.8\% \\
5  &           \textcolor{blue}{WhoFramedRogerRabbit} &    491 &                \textcolor{blue}{ReturnOfTheJedi} &   1926 &     +519.3\% \\
6  &                       \textcolor{blue}{Serenity} &    472 &                 \textcolor{blue}{XMenFilmSeries} &   1870 &     +307.4\% \\
7  &                   \textcolor{blue}{Transformers} &    459 &                      \textcolor{blue}{TheMatrix} &   1861 &     +169.3\% \\
8  &                           \textcolor{blue}{XMen} &    459 &                   \textcolor{blue}{StarTrek2009} &   1769 &     +214.8\% \\
9  &                 \textcolor{blue}{XMenFirstClass} &    446 &                                  AvengersEndgame &   1765 &           -- \\
10 &                                     Ghostbusters &    445 &                              AvengersInfinityWar &   1729 &  +28,716.7\% \\
11 &                     \textcolor{blue}{MenInBlack} &    427 &                                  TheForceAwakens &   1717 &   +9,438.9\% \\
12 &                         \textcolor{blue}{Avatar} &    426 &                                 RevengeOfTheSith &   1706 &     +635.3\% \\
13 &             \textcolor{blue}{TheDarkKnightRises} &    408 &                           \textcolor{blue}{Thor} &   1697 &     +366.2\% \\
14 &           \textcolor{blue}{XMenDaysOfFuturePast} &    394 &                                     ThorRagnarok &   1640 &           -- \\
15 &                       \textcolor{blue}{KillBill} &    381 &                              AvengersAgeOfUltron &   1585 &           -- \\
16 &                           \textcolor{blue}{Thor} &    364 &           \textcolor{blue}{TheEmpireStrikesBack} &   1584 &     +504.6\% \\
17 &                \textcolor{blue}{BackToTheFuture} &    360 &                   CaptainAmericaTheWinterSoldier &   1483 &     +493.2\% \\
18 &                                       Spaceballs &    359 &                           CaptainAmericaCivilWar &   1481 &  +21,057.1\% \\
19 &                     \textcolor{blue}{PacificRim} &    356 &             \textcolor{blue}{TheDarkKnightRises} &   1452 &     +255.9\% \\
20 &                    \textcolor{blue}{HarryPotter} &    346 &           \textcolor{blue}{WhoFramedRogerRabbit} &   1444 &     +194.1\% \\
21 &     \textcolor{blue}{MontyPythonAndTheHolyGrail} &    345 &                  \textcolor{blue}{TheWizardOfOz} &   1437 &     +380.6\% \\
22 &                                 ThePrincessBride &    342 &           \textcolor{blue}{XMenDaysOfFuturePast} &   1425 &     +261.7\% \\
23 &                   \textcolor{blue}{BatmanBegins} &    339 &  \textcolor{blue}{CaptainAmericaTheFirstAvenger} &   1420 &     +346.5\% \\
24 &                    \textcolor{blue}{TheAvengers} &    337 &                 \textcolor{blue}{XMenFirstClass} &   1420 &     +218.4\% \\
25 &                                  TheThreeStooges &    336 &                                      TheLastJedi &   1397 &           -- \\
26 &                                          IronMan &    327 &                                 ThePhantomMenace &   1379 &     +529.7\% \\
27 &  \textcolor{blue}{CaptainAmericaTheFirstAvenger} &    318 &           \textcolor{blue}{GuardiansOfTheGalaxy} &   1373 &     +406.6\% \\
28 &                         \textcolor{blue}{AlienS} &    314 &                       \textcolor{blue}{Serenity} &   1355 &     +187.1\% \\
29 &                \textcolor{blue}{ReturnOfTheJedi} &    311 &         \textcolor{blue}{Terminator2JudgmentDay} &   1352 &     +349.2\% \\
30 &         \textcolor{blue}{Terminator2JudgmentDay} &    301 &                   \textcolor{blue}{Transformers} &   1231 &     +168.2\% \\
31 &                  \textcolor{blue}{TheWizardOfOz} &    299 &                         \textcolor{blue}{Aliens} &   1229 &     +291.4\% \\
32 &                                          HotFuzz &    296 &                       \textcolor{blue}{KillBill} &   1226 &     +221.8\% \\
33 &       \textcolor{blue}{StarTrekIITheWrathOfKhan} &    295 &                                   XMenApocalypse &   1226 &   +8,073.3\% \\
34 &                                  IndependenceDay &    288 &                \textcolor{blue}{BackToTheFuture} &   1219 &     +238.6\% \\
35 &           \textcolor{blue}{StarTrekIntoDarkness} &    283 &       \textcolor{blue}{StarTrekIITheWrathOfKhan} &   1219 &     +313.2\% \\
36 &                                  TheFifthElement &    276 &           \textcolor{blue}{StarTrekIntoDarkness} &   1168 &     +312.7\% \\
37 &                                      GalaxyQuest &    276 &                         \textcolor{blue}{Avatar} &   1167 &     +173.9\% \\
38 &                     \textcolor{blue}{ManOfSteel} &    274 &                     \textcolor{blue}{PacificRim} &   1165 &     +227.2\% \\
39 &                                       TronLegacy &    271 &                     BatmanVSupermanDawnOfJustice &   1164 &  +11,540.0\% \\
40 &           \textcolor{blue}{GuardiansOfTheGalaxy} &    271 &                     \textcolor{blue}{Batman1989} &   1161 &     +333.2\% \\
41 &                                     AustinPowers &    268 &                    \textcolor{blue}{HarryPotter} &   1146 &     +231.2\% \\
42 &                                          DieHard &    268 &                                 BlackPanther2018 &   1126 &           -- \\
43 &                         \textcolor{blue}{Batman} &    268 &                     \textcolor{blue}{ManOfSteel} &   1124 &     +310.2\% \\
44 &                   \textcolor{blue}{TheGodfather} &    267 &                   \textcolor{blue}{TheGodfather} &   1122 &     +320.2\% \\
45 &                                        Inception &    267 &     \textcolor{blue}{MontyPythonAndTheHolyGrail} &   1117 &     +223.8\% \\
46 &                                         IronMan2 &    265 &                                AttackOfTheClones &   1113 &     +525.3\% \\
47 &           \textcolor{blue}{TheEmpireStrikesBack} &    262 &                     \textcolor{blue}{MenInBlack} &   1105 &     +158.8\% \\
48 &                                     Godzilla2014 &    257 &                         GuardiansOfTheGalaxyVol2 &   1104 &           -- \\
49 &                              GIJoeTheRiseofCobra &    257 &                   \textcolor{blue}{BatmanBegins} &   1103 &     +225.4\% \\
\bottomrule
\end{tabular}

\caption{Comparative top fifty films by the number of tropes between July 2016 and April 2019. *Common elements are marked in blue.}
\label{tab:top_films_by_number_of_tropes}
\end{table}

In some cases, the films have changed the name from 2016 to avoid ambigüity
(TheAvengers becomes TheAvengers2012, StarTrek becomes StarTrek2009, AlienS becomes Aliens,
Batman becomes Batman1989, Xmen becomes XmenFilmSeries).
As shown in the column "Increment", most of the films have around 100\%-300\%
more tropes in 2020 than in 2016.
In just some cases, some films have increased the number of tropes drastically,
to cite the ones with an increment above 500\%: TheAvengers, ANewHope, ReturnOfTheJedi, RevengeOfTheSith,
TheEmpireStrikesBack, ThePhantomMenace and AttackOfTheClones.
It is interesting that 6 out of 7 films with such increment are from the Star Wars franchise,
that, in 2016 (with Episode VI just released), only had two films in the top and, in 2020,
with just four films more released by then, has seven.

Thirty-four films out of the top fifty were already in the top list in 2016.
The remaining set includes films that had just been released at that time
and had a limited number of tropes (CaptainAmericaCivilWar, TheForceAwakens, XMenApocalyse,
BatmanVSupermanDawnOfJustice).
Also, films that were released but not included in TV Tropes yes (AvengersAgeOfUltron)
and films not released at that time (AvengerEndGame, ThorRagnarok, BlackPanther2018, GuardiansOfTheGalaxyVol2).
Finally, TV Tropes also includes films that are not released yet but are already fulfilling some meta-tropes;
this is the case of AvengersInfinityWar, that already had the tropes LoadsAndLoadsOfCharacters
and GrandFinale in 2016.

\section{Analysis of the number of films where the tropes appear} \label{sec:analysys2}

\begin{figure}[htpb]
\center

\includegraphics[width= 10cm]{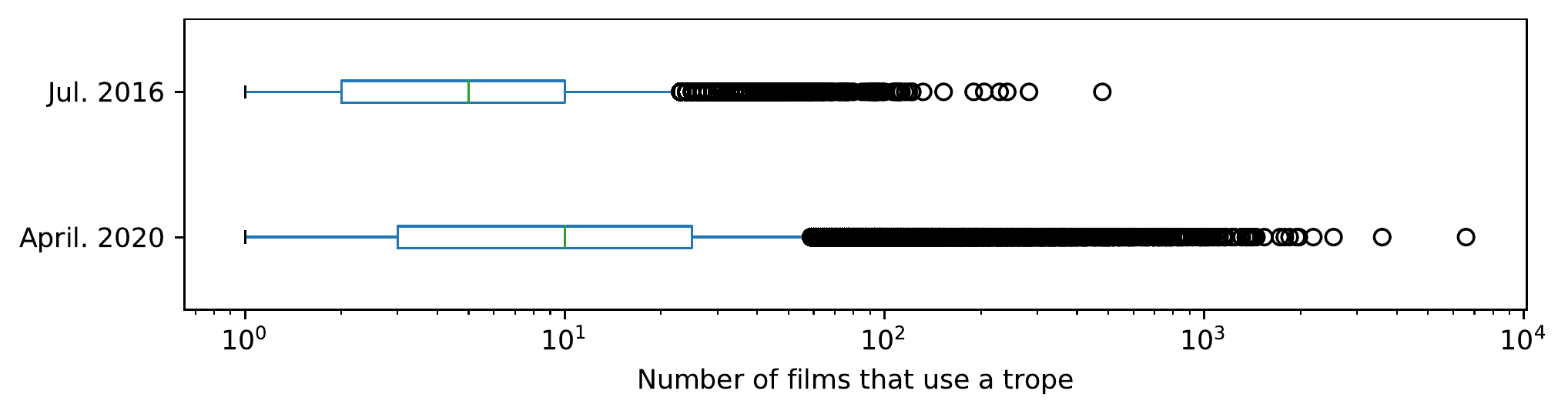}

\caption{Comparative boxplots of trope occurrences by films between July 2016 and April 2019}
\label{fig:boxplot_films_by_trope}
\end{figure}

The descriptive analysis of the tropes and the films they appear
in (Table~\ref{tab:decriptive_analysis_films_by_trope})
reveals a growth of the number of tropes by 110.38\%.
The average number of films that every trope appears in also grows by 259.95\%.
While there are tropes that only exist in a single film, every trope is used on average by 28 films.
The comparative boxplots (Figure~\ref{fig:boxplot_films_by_trope}) shows a more significant
interquartile range, and a shift to the top,
that corresponds to stretching the histogram from 2016 and a longer tail.

\begin{table}[htpb]
\center

\begin{tabular}{lrr}
\toprule
{} &  Jul. 2016 &  April. 2020 \\
\midrule
min      &      1.000 &        1.000 \\
max      &    480.000 &     6591.000 \\
nobs     &  17738.000 &    37317.000 \\
mean     &      7.803 &       28.087 \\
kurtosis &    274.065 &     1186.002 \\
skewness &     10.043 &       22.339 \\
variance &    117.915 &     7111.168 \\
\bottomrule
\end{tabular}

\caption{Comparative descriptive analysis of trope occurrences by film between July 2016 and April 2019}
\label{tab:decriptive_analysis_films_by_trope}
\end{table}

\begin{figure}[htpb]
\center

\includegraphics[width= 10cm]{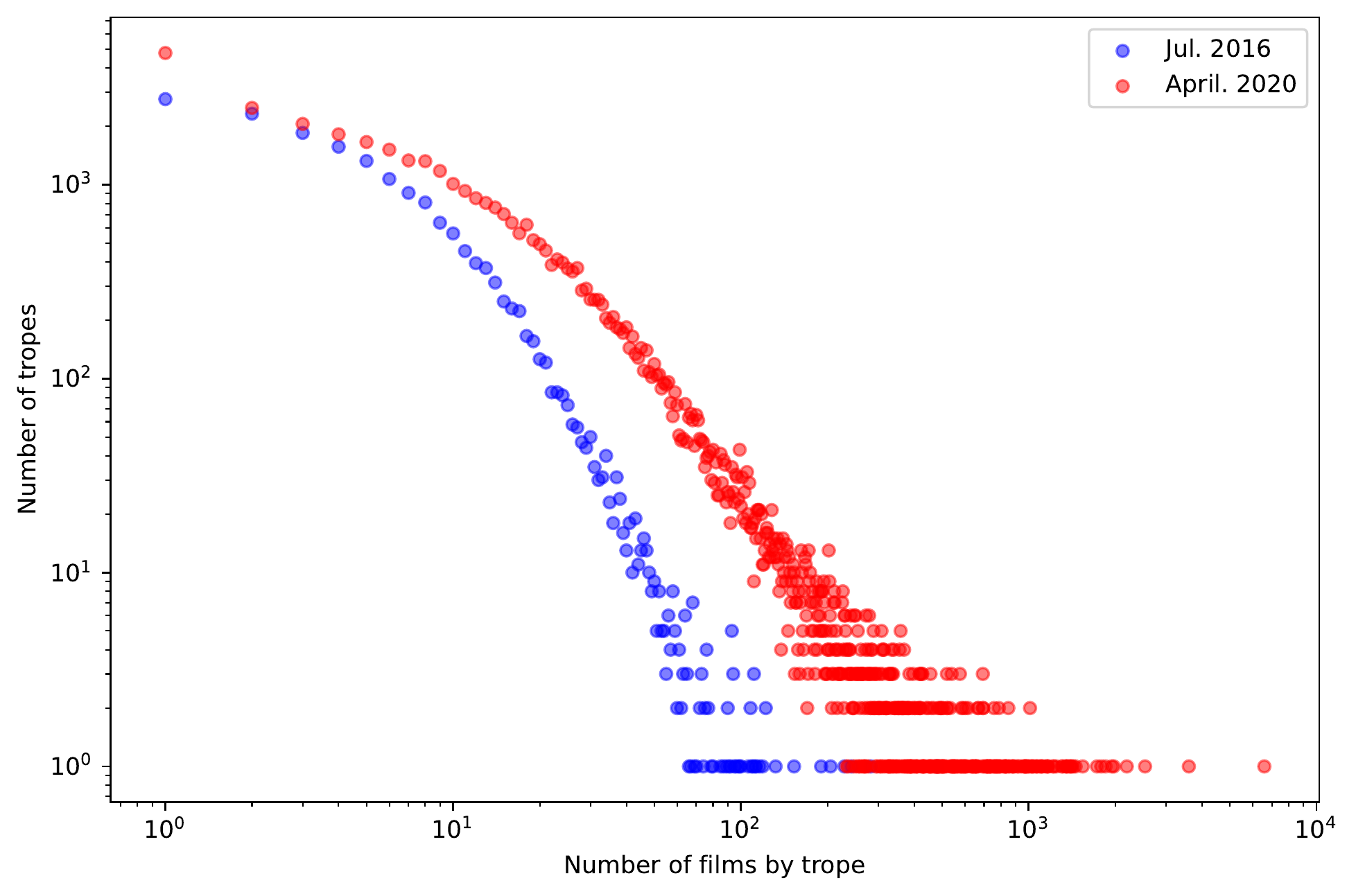}

\caption{Comparative boxplots of trope occurrences by film between July 2016 and April 2019}
\label{fig:frequency_films_by_trope}
\end{figure}

The representation of the number of films vs tropes (Figure~\ref{fig:frequency_films_by_trope})
shows a distribution similar to 2016 but stretched and extended.
The only main difference is that the number of films with a single trope
is three times as high now as in 2016.
In other words, we have more tropes, and in general, films have more tropes as well.

However, having a distribution that resembles the one in 2016
does not necessarily mean that TV Tropes has not suffered many changes.
Although the top films have not changed a lot since 2016, the top tropes have,
keeping only three out of fifty (Table~\ref{tab:top_tropes_by_number_of_films}).
The tropes have evolved so much since 2016 that 43 out of the top fifty in 2020
did not even exist in 2016.

\begin{table}[htpb]
\center

\begin{tabular}{llrlrl}
\toprule
{} &                           Trope (Jul. 2016) & Films &                         Trope (April. 2020) & Films &     Increment \\
\midrule
0  &             \textcolor{blue}{BoxOfficeBomb} &   480 &                               AmericanFilms &  6591 &            -- \\
1  &                                 CultClassic &   283 &                                    ShoutOut &  3603 &            -- \\
2  &                               HeyItsThatGuy &   242 &                                      BigBad &  2538 &            -- \\
3  &                                   ArchEnemy &   229 &             \textcolor{blue}{BoxOfficeBomb} &  2193 &      +356.9\% \\
4  &                                      BMovie &   205 &                                 ChekhovsGun &  1977 &            -- \\
5  &                          CaliforniaDoubling &   190 &                                 HorrorFilms &  1944 &            -- \\
6  &                                    LogoJoke &   153 &                               Foreshadowing &  1850 &  +184,900.0\% \\
7  &          HorrorDoesntSettleForSimpleTuesday &   132 &                                      OhCrap &  1787 &            -- \\
8  &                      RetroactiveRecognition &   122 &                           BittersweetEnding &  1727 &            -- \\
9  &                               SignatureSong &   122 &                                   TitleDrop &  1541 &            -- \\
10 &                                    FilmNoir &   119 &                              DeadpanSnarker &  1456 &            -- \\
11 &                          DisneyVillainDeath &   116 &              \textcolor{blue}{OneWordTitle} &  1429 &    +1,329.0\% \\
12 &                                DuringTheWar &   113 &                             FilmsOfThe1980s &  1410 &            -- \\
13 &                            ProtagonistTitle &   112 &                                DownerEnding &  1400 &            -- \\
14 &                           PreMortemOneLiner &   111 &                                         WMG &  1396 &            -- \\
15 &  \textcolor{blue}{CompletelyDifferentTitle} &   111 &                                     Jerkass &  1377 &            -- \\
16 &                              RomanticComedy &   111 &                               NightmareFuel &  1356 &            -- \\
17 &                         BillingDisplacement &   110 &                                    LargeHam &  1353 &            -- \\
18 &                          TheForeignSubtitle &   109 &                             FilmsOf20102014 &  1349 &            -- \\
19 &                         MissingTrailerScene &   108 &                             FilmsOf20052009 &  1311 &            -- \\
20 &                             FakeNationality &   108 &                                  TearJerker &  1307 &            -- \\
21 &                         HollywoodActionHero &   106 &                                  RunningGag &  1247 &            -- \\
22 &              \textcolor{blue}{OneWordTitle} &   100 &                              Headscratchers &  1221 &            -- \\
23 &                                   EpicMovie &    99 &                      WhatHappenedToTheMouse &  1217 &            -- \\
24 &                                  TheStinger &    98 &                             FilmsOf20152019 &  1167 &            -- \\
25 &                              DemotedToExtra &    96 &                                KarmaHoudini &  1163 &            -- \\
26 &                 ImpaledWithExtremePrejudice &    95 &                                    TheCameo &  1162 &    +2,012.7\% \\
27 &                                PrettyInMink &    94 &                              MeaningfulName &  1146 &            -- \\
28 &                           SpiritualLicensee &    94 &                               AssholeVictim &  1116 &            -- \\
29 &                                 AllStarCast &    94 &                             DrivenToSuicide &  1111 &            -- \\
30 &                   BestKnownForTheFanservice &    93 &                                     Tagline &  1105 &            -- \\
31 &                                 AndStarring &    93 &                               TooDumbToLive &  1096 &            -- \\
32 &                          CriticalDissonance &    93 &                 MohsScaleOfViolenceHardness &  1071 &            -- \\
33 &                                CreatorCameo &    93 &                                 ImageSource &  1067 &            -- \\
34 &                                   FinalGirl &    93 &                               ActorAllusion &  1061 &            -- \\
35 &                                  SequelHook &    92 &                                  KickTheDog &  1039 &            -- \\
36 &                         BottomlessMagazines &    91 &                                    BookEnds &  1034 &            -- \\
37 &                               SignatureLine &    90 &                               BerserkButton &  1025 &            -- \\
38 &                               DVDCommentary &    90 &                             FilmsOf20002004 &  1022 &            -- \\
39 &               NotEvenBotheringWithTheAccent &    89 &                           YourCheatingHeart &  1011 &    +2,251.2\% \\
40 &                                     TheOner &    87 &                               RealityEnsues &  1011 &  +101,000.0\% \\
41 &                                     Fingore &    85 &                            PrecisionFStrike &  1002 &            -- \\
42 &                                       CarFu &    80 &                                MoodWhiplash &   997 &            -- \\
43 &                                 ThreeDMovie &    79 &  \textcolor{blue}{CompletelyDifferentTitle} &   984 &      +786.5\% \\
44 &                      DeliberatelyMonochrome &    77 &                                 GroinAttack &   983 &            -- \\
45 &                                    NeckSnap &    77 &                                   BrickJoke &   979 &            -- \\
46 &                          BlackDudeDiesFirst &    76 &                             FilmsOfThe1970s &   973 &            -- \\
47 &                           DeathByAdaptation &    76 &                 EstablishingCharacterMoment &   969 &            -- \\
48 &                                    ArcWords &    76 &                    FilmsDiscussedByMoviebob &   967 &            -- \\
49 &                    ArtisticLicenseGeography &    76 &                                BritishFilms &   964 &            -- \\
\bottomrule
\end{tabular}

\caption{Comparative top fifty films by the number of tropes between July 2016 and April 2019. *Common elements are marked in blue.}
\label{tab:top_tropes_by_number_of_films}
\end{table}

The first question that comes into our minds is if the scraper is working fine.
In order to answer this question, we analyse the top tropes from 2016
and try to trace the reasons why they are not so popular in 2020.
Table~\ref{tab:old_top_tropes_and_increment} shows
the comparative ranking of the top tropes between 2016 and their position in 2020.
According to it, the vast majority of tropes have incremented their appearance in films by a 200\%-300\%.
Nevertheless, this is not enough for them to be in the top fifty,
although they mostly keep being in the top 1,300.
There are four cases with a reduced use: TV Tropes avoids \textit{CultClassic} now because of its lack of universality,
removes \textit{HeyItsThatGuy} and replaces it by \textit{RetroactiveRecognition},
replaces \textit{SpiritualLicensee} by \textit{SpiritualAdaptation} (although the old trope still appears in some films)
and removes \textit{ThreeDMovie}.

\begin{table}[htpb]
\center

\begin{tabular}{llrrll}
\toprule
{} &                               Trope & Films (Jul. 2016) & Films (April. 2020) &   Increment &                     Moves to \\
\midrule
0  &                       BoxOfficeBomb &               480 &                2193 &    +356.9\% &       +3\textsuperscript{rd} \\
1  &                         CultClassic &               283 &                  13 &     -95.4\% &  +16,030\textsuperscript{th} \\
2  &                       HeyItsThatGuy &               242 &                   0 &          -- &                           -- \\
3  &                           ArchEnemy &               229 &                 447 &     +95.2\% &     +212\textsuperscript{th} \\
4  &                              BMovie &               205 &                 577 &    +181.5\% &     +138\textsuperscript{th} \\
5  &                  CaliforniaDoubling &               190 &                 418 &    +120.0\% &     +238\textsuperscript{th} \\
6  &                            LogoJoke &               153 &                 420 &    +174.5\% &     +231\textsuperscript{st} \\
7  &  HorrorDoesntSettleForSimpleTuesday &               132 &                 179 &     +35.6\% &     +854\textsuperscript{th} \\
8  &              RetroactiveRecognition &               122 &                 505 &    +313.9\% &     +170\textsuperscript{th} \\
9  &                       SignatureSong &               122 &                 319 &    +161.5\% &     +370\textsuperscript{th} \\
10 &                            FilmNoir &               119 &                 190 &     +59.7\% &     +791\textsuperscript{st} \\
11 &                  DisneyVillainDeath &               116 &                 400 &    +244.8\% &     +251\textsuperscript{st} \\
12 &                        DuringTheWar &               113 &                 130 &     +15.0\% &   +1,292\textsuperscript{nd} \\
13 &                    ProtagonistTitle &               112 &                 204 &     +82.1\% &     +705\textsuperscript{th} \\
14 &                   PreMortemOneLiner &               111 &                 385 &    +246.8\% &     +266\textsuperscript{th} \\
15 &            CompletelyDifferentTitle &               111 &                 984 &    +786.5\% &      +43\textsuperscript{rd} \\
16 &                      RomanticComedy &               111 &                 314 &    +182.9\% &     +376\textsuperscript{th} \\
17 &                 BillingDisplacement &               110 &                 241 &    +119.1\% &     +559\textsuperscript{th} \\
18 &                  TheForeignSubtitle &               109 &                 208 &     +90.8\% &     +689\textsuperscript{th} \\
19 &                 MissingTrailerScene &               108 &                 266 &    +146.3\% &     +490\textsuperscript{th} \\
20 &                     FakeNationality &               108 &                 276 &    +155.6\% &     +468\textsuperscript{th} \\
21 &                 HollywoodActionHero &               106 &                 136 &     +28.3\% &   +1,211\textsuperscript{th} \\
22 &                        OneWordTitle &               100 &                1429 &  +1,329.0\% &      +11\textsuperscript{th} \\
23 &                           EpicMovie &                99 &                 219 &    +121.2\% &     +639\textsuperscript{th} \\
24 &                          TheStinger &                98 &                 456 &    +365.3\% &     +205\textsuperscript{th} \\
25 &                      DemotedToExtra &                96 &                 412 &    +329.2\% &     +241\textsuperscript{st} \\
26 &         ImpaledWithExtremePrejudice &                95 &                 484 &    +409.5\% &     +188\textsuperscript{th} \\
27 &                        PrettyInMink &                94 &                 303 &    +222.3\% &     +405\textsuperscript{th} \\
28 &                   SpiritualLicensee &                94 &                  35 &     -62.8\% &   +6,670\textsuperscript{th} \\
29 &                         AllStarCast &                94 &                 178 &     +89.4\% &     +861\textsuperscript{st} \\
30 &           BestKnownForTheFanservice &                93 &                 226 &    +143.0\% &     +613\textsuperscript{th} \\
31 &                         AndStarring &                93 &                 437 &    +369.9\% &     +216\textsuperscript{th} \\
32 &                  CriticalDissonance &                93 &                 272 &    +192.5\% &     +478\textsuperscript{th} \\
33 &                        CreatorCameo &                93 &                 765 &    +722.6\% &      +77\textsuperscript{th} \\
34 &                           FinalGirl &                93 &                 373 &    +301.1\% &     +281\textsuperscript{st} \\
35 &                          SequelHook &                92 &                 517 &    +462.0\% &     +165\textsuperscript{th} \\
36 &                 BottomlessMagazines &                91 &                 295 &    +224.2\% &     +416\textsuperscript{th} \\
37 &                       SignatureLine &                90 &                 267 &    +196.7\% &     +489\textsuperscript{th} \\
38 &                       DVDCommentary &                90 &                 198 &    +120.0\% &     +742\textsuperscript{nd} \\
39 &       NotEvenBotheringWithTheAccent &                89 &                 320 &    +259.6\% &     +367\textsuperscript{th} \\
40 &                             TheOner &                87 &                 340 &    +290.8\% &     +332\textsuperscript{nd} \\
41 &                             Fingore &                85 &                 363 &    +327.1\% &     +296\textsuperscript{th} \\
42 &                               CarFu &                80 &                 349 &    +336.2\% &     +321\textsuperscript{st} \\
43 &                         ThreeDMovie &                79 &                   1 &     -98.7\% &  +35,034\textsuperscript{th} \\
44 &              DeliberatelyMonochrome &                77 &                 313 &    +306.5\% &     +381\textsuperscript{st} \\
45 &                            NeckSnap &                77 &                 254 &    +229.9\% &     +524\textsuperscript{th} \\
46 &                  BlackDudeDiesFirst &                76 &                 402 &    +428.9\% &     +247\textsuperscript{th} \\
47 &                   DeathByAdaptation &                76 &                 397 &    +422.4\% &     +252\textsuperscript{nd} \\
48 &                            ArcWords &                76 &                 639 &    +740.8\% &     +116\textsuperscript{th} \\
49 &            ArtisticLicenseGeography &                76 &                 262 &    +244.7\% &     +501\textsuperscript{st} \\
\bottomrule
\end{tabular}

\caption{Top fifty films by the number of tropes in July 2016 and their increment of tropes in April 2020.}
\label{tab:old_top_tropes_and_increment}
\end{table}

\section{Conclusions} \label{sec:conclusions}

We extract the following outcomes from the previous analysis.

In terms of gross numbers, the number of films and tropes in 2020 doubles the number in 2016.
It goes from 6,296 films to 12,567 and from 17,738 tropes to 37,317.
The average number of relations between films and tropes in 2020 quadruples the number in 2016;
in other words, films are better described in terms of tropes and tropes appear in more films.
In absolute terms, the number of connections has increased by 657.23\%.
Both distribution curves shift right (less positive skewness).
However, while the interquartile range of tropes' distribution has reduced,
the interquartile range of the films' distribution has increased.

Regarding the films with more tropes in TV Tropes,
only 32\% of the films in the top fifty have changed from 2016 to 2020,
that is to say, in general, popular films in terms of tropes remain.
The James Bond franchise is the artistic work with more tropes,
with almost a third more than the following, The Lord of the rings,
that overcomes the next by a few hundreds of tropes, The Avengers.
It is imperative to hint how the franchise Star Wars boosted its popularity from 2016 to 2020, also.
Regarding the most popular tropes, although the majority of them in 2016 are still relevant in 2020,
the new top list if almost entirely replaced by tropes that did not exist in 2016.
The most popular trope in 2020 is \textit{AmericanFilms}, followed by \textit{ShoutOut} and \textit{BigBad}.

We discovered that TV Tropes adapts the name of the films to disambiguate second parts or remakes as they appear.
They also include films that have not been premiered and extrapolate possible tropes from the existing information.

We could detect errors in the wiki that lead to parsing inexisting film pages
and incomplete information that made us evolve the scraper several times.
The latest modifications made to tropescraper allows extracting the data correctly,
with results that correlate to the ones from DB Tropes in 2016.
However, due that the wiki is unstructured and maintained manually,
it is crucial to keep a critic eye on the data extracted from it, especially the outliers,
and always consider a data wrangling phase before using the data to build models in machine learning.
It also points to specific, and network-based, dynamic of tropes and films.

\section*{Acknowledgements}
This work has been partially funded by projects DeepBio (TIN2017-85727-C4-2-P) and TEC2015-68752
and ``Ayuda del Programa de Fomento e Impulso de la actividad Investigadora de la Universidad de C\'adiz''.

\section{Bibliography}
\bibliographystyle{abbrv}
\bibliography{report}

\end{document}